\documentclass[preliminary,copyright,noncommercial]{eptcs}
\providecommand{\event}{1st Graphs as Models (GAM) Workshop, ETAPS-2015}
\usepackage{amsmath}
\usepackage{graphicx}
\usepackage[english]{babel}
\usepackage{hyperref}
\usepackage{float}
\usepackage[justification=centering]{caption}
\newcommand{\squeezeup}{\vspace{-2mm}}
\title{Detecting and Refactoring Operational Smells within the Domain Name System}
\author{Marwan Radwan \qquad\qquad Reiko Heckel
\institute{Computer Science Department\\
University of Leicester\\
Leicester - United Kingdom}
\email{mmmr1@mcs.le.ac.uk \qquad\qquad\qquad reiko@mcs.le.ac.uk}
}
\def\titlerunning{Detecting and Refactoring Operational Smells within the Domain Name System}
\def\authorrunning{M. Radwan \& R. Heckel}

\begin{document}

\maketitle
\squeezeup
 \squeezeup
\begin{abstract}
The Domain Name System (DNS) is one of the most important components of the Internet infrastructure. DNS relies on a delegation-based architecture, where resolution of names to their IP addresses requires resolving the names of the servers responsible for those names.  The recursive structures of the inter-dependencies that exist between name servers associated with each zone are called dependency graphs. System administrators' operational decisions have far reaching effects on the DNS’s qualities. They need to be soundly made to create a balance between the availability, security and resilience of the system. We utilize dependency graphs to identify, detect and catalogue \textit {operational bad smells}. Our method deals with smells on a high-level of abstraction using a consistent taxonomy and reusable vocabulary, defined by a \textit{DNS Operational Model}. The method will be used to build a diagnostic advisory tool that will detect configuration changes that might decrease the robustness or security posture of domain names before they become into production.
\end{abstract}
\squeezeup
 \squeezeup
  \squeezeup
\section{Introduction}
\label{sec:sec1}
The Domain Name System (DNS) \cite{rfc1034} is critical to the integrity of Internet services and applications. The DNS is a distributed database for storing information on domain names, the primary namespace for hosts on the Internet. The name space is organised in a hierarchical structure to ensure domain name uniqueness. Each node in the DNS tree corresponds to a zone. Each zone belonging to a single administrative authority is served by multiple authoritative name servers. \par

The correct and error-free operation of the DNS is crucial for the reliability of most applications on the Internet. Operational guidelines \cite{rfc1912,rfc2182,wg4} require that a zone have multiple authoritative name servers, and that they be distributed through diverse network topological and geographical locations to increase the reliability of that zone as well as improve overall network performance and access. It also makes DNS services robust against unexpected failures. Recent work \cite{deccio2010,osterweil2011} outlines the need for zone operators to understand how many inter-dependencies they may inadvertently be incurring through the deployment and sharing of DNS secondary servers. \par

The original DNS design focused mainly on system robustness against physical failures, and neglected the impact of operational errors such as misconfigurations and bad deployment choices. Several previous measurements \cite{pappas2009,kalafut2008,wessels2004} showed that zones with configuration errors suffer from reduced availability and increased query delays up to an order of magnitude. DNS administrators have to decide on operational parameters and be aware of their implications for the DNS's overall system qualities. On the deployment level, configuring the number of redundant authoritative DNS servers for a certain zone shall take into consideration the operational overhead associated with querying multiple servers in parallel. Choosing servers with names under other zones provides zone redundancy but may incur security and resiliency threats to the zone. Deciding on where to physically locate the servers should ensure a certain degree of resistance against different types of failures. Peering with external organizations for secondary server hosting should take into consideration the impact of transitional trust and administrative complexity \cite{ram2005,herzberg2013}.\par

 While the original DNS design documents \cite{rfc1033,rfc1034,rfc1035,rfc1912,rfc2182} call for diverse placement of authoritative name servers for a zone, bad configurations may lead to \textit {cyclic dependencies} while bad deployment choices may lead to \textit{diminished and false server redundancy}. It was also assumed that redundant DNS servers fail independently; previous measurements \cite{ram2005,deccio2010} showed that operational deployment choices made at individual zones can introduce \textit {excessive zone influence} that severely affect the availability, security and resiliency of other zones. \par

This research is motivated by the lack of formal analysis of the DNS interdependencies stemming from the delegation-based architecture as well as operational deployment choices made by system administrators. We approached the problem from a design point of view that takes into consideration the DNS zone configuration and server deployment choices rather than from the dynamic behavioural view \cite{casalicchio2012} which includes statistical and post-deployment measurements. We propose a method to identify, specify and detect misconfigurations and bad deployment choices in the form of operational bad smells. \par

The method utilizes a set of structural metrics defined over a DNS operational model to detect the smells in early stages of the DNS deployment. It also suggests graph-based refactoring rules as correction mechanisms for the bad smells. We apply and validate the method using several representative case studies. The method will be used to build a \textit {pre-emptive diagnostic advisory} tool that will detect and flag configuration changes that might decrease the robustness or security posture of a domain name, before even the changes become into production. The contributions of this research are:
\begin{enumerate}

\item	Introduction of the concept of operational bad smells, i.e., recurring DNS deployment bad choices and misconfigurations that have negative impact on certain aspects of the DNS's quality.
\item	Description in detail of a set of representative operational bad smells to build a DNS operational bad smells catalogue.
\item	Identification of a set of structural metrics, defined over a DNS operational model, to query the dependency graph of the system to detect DNS operational bad smells.
\item	Suggestion of graph-based refactoring rules as correction mechanisms to eliminate the bad smells.
\end{enumerate}

The rest of the paper is structured as follows: Section~\ref{sec:sec2} discusses relevant background. Section~\ref{sec:sec3} presents the DNS operational model. Section~\ref{sec:sec4} discusses the bad smells' identification, specification, detection and refactoring method. Section~\ref{sec:sec5} validates our method by applying it to a set of representative case studies. Section~\ref{sec:sec6} discusses some related work and Section~\ref{sec:sec7} concludes the paper and discusses future work.
\squeezeup
\squeezeup
\section{The Operation and Structure of the DNS}
\label{sec:sec2}
DNS is responsible for the mapping of human-friendly domain names to the corresponding machine-oriented IP addresses. Operators of each zone determine the number of authoritative name servers and their placement and manage all changes to the zone's data content. In spite of the fact that zone administration is autonomous, some coordination is required to maintain the consistency of the DNS hierarchy.
 \squeezeup
 \squeezeup
\subsection{General Operation of the DNS}
\begin{figure} [ht]
  \centering
  \includegraphics[trim=0cm 1cm 0cm 0.5cm, clip=true,width=0.75\textwidth]{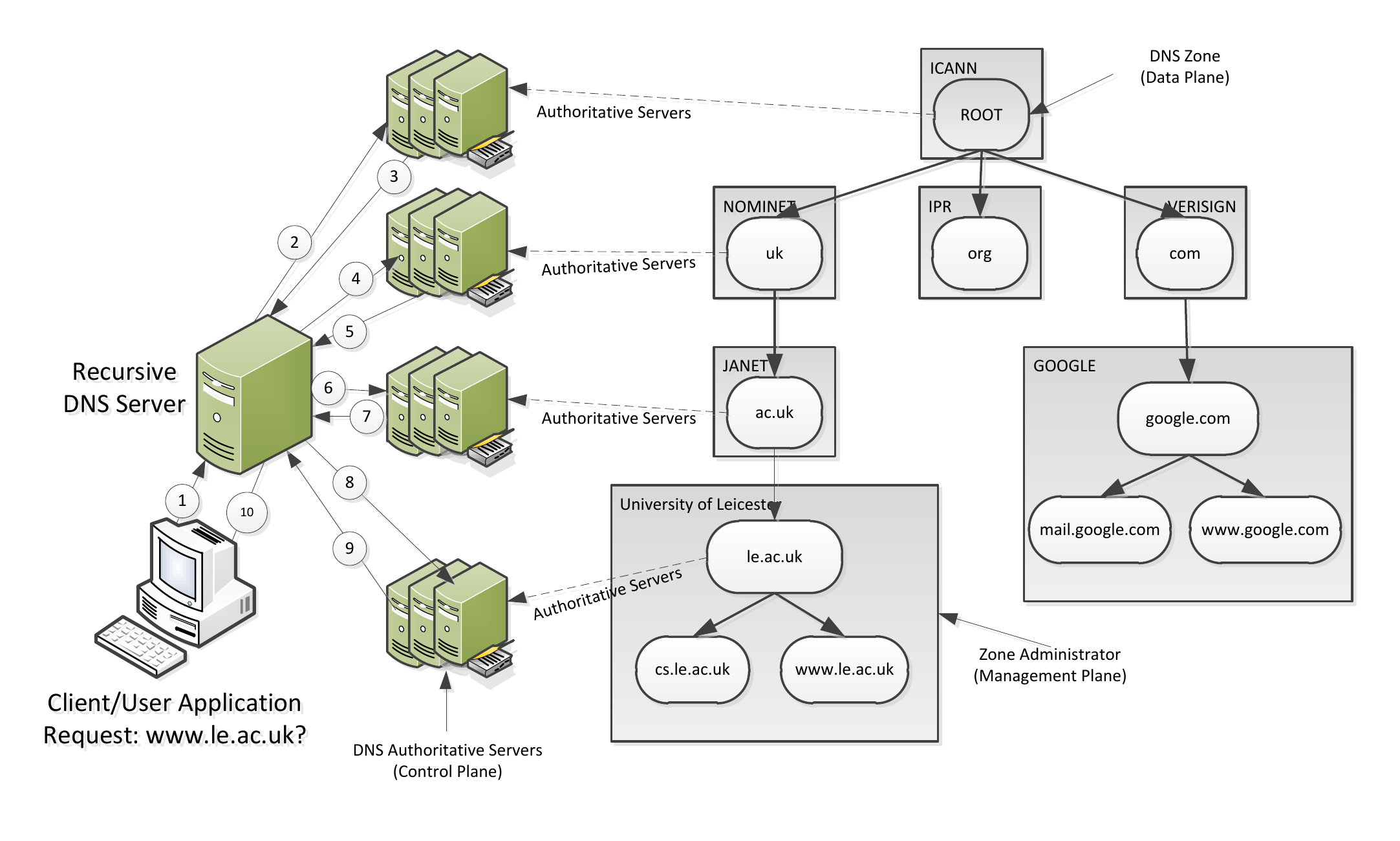}
 \caption{An illustration of the DNS resolution process.}
 \label{fig:f1}
  \squeezeup
   \squeezeup
\end{figure}

Figure~\ref{fig:f1} shows the process by which an application looks up the domain name www.le.ac.uk and how it is mapped to the DNS data, control and management operational planes.
To find the IP address of www.le.ac.uk, the client (e.g a web browser) submits a DNS query to a recursive DNS resolver (step 1). Assuming that the corresponding IP is not in the resolver cache, it will ask one of the root name servers for the translation (step 2). The names and IP addresses of root name servers are locally stored within each server. The root name servers will respond with a “referral”, telling the resolver to query the DNS servers of the .uk domain for an answer (step 3). The resolver then repeats this process for the .uk name servers and get a referral to ask the .ac.uk name servers which in turn answers with a referral to as the le.ac.uk name servers (step 4 -7). The resolver next asks one of the le.ac.uk name servers for the translation (step 8), and gets the answer in step (9), and finally forwards the answer to the requesting client (step 10) who will use this information to connect to the web server hosting the web site www.le.ac.uk. Throughout the process, resolvers may encounter name servers hosted under other zones whose names need to be resolved before contacting them about the original request.
\squeezeup
\squeezeup
\subsection{DNS Operational Inter-dependencies}

Inter-dependencies are common in the DNS and stem from the hierarchal structure of the DNS, the DNS protocol as well as from different motivations and goals \cite{deccio2010}. A zone is said to depend on a name server if the name server could be involved in the resolution of names in that zone. The dependencies among name servers that directly or indirectly affect a zone are represented as a dependency graph.\par
\begin{figure}
  \centering
  \includegraphics[ trim=0cm 0cm 0cm 0cm, clip=true,width=0.75\textwidth]{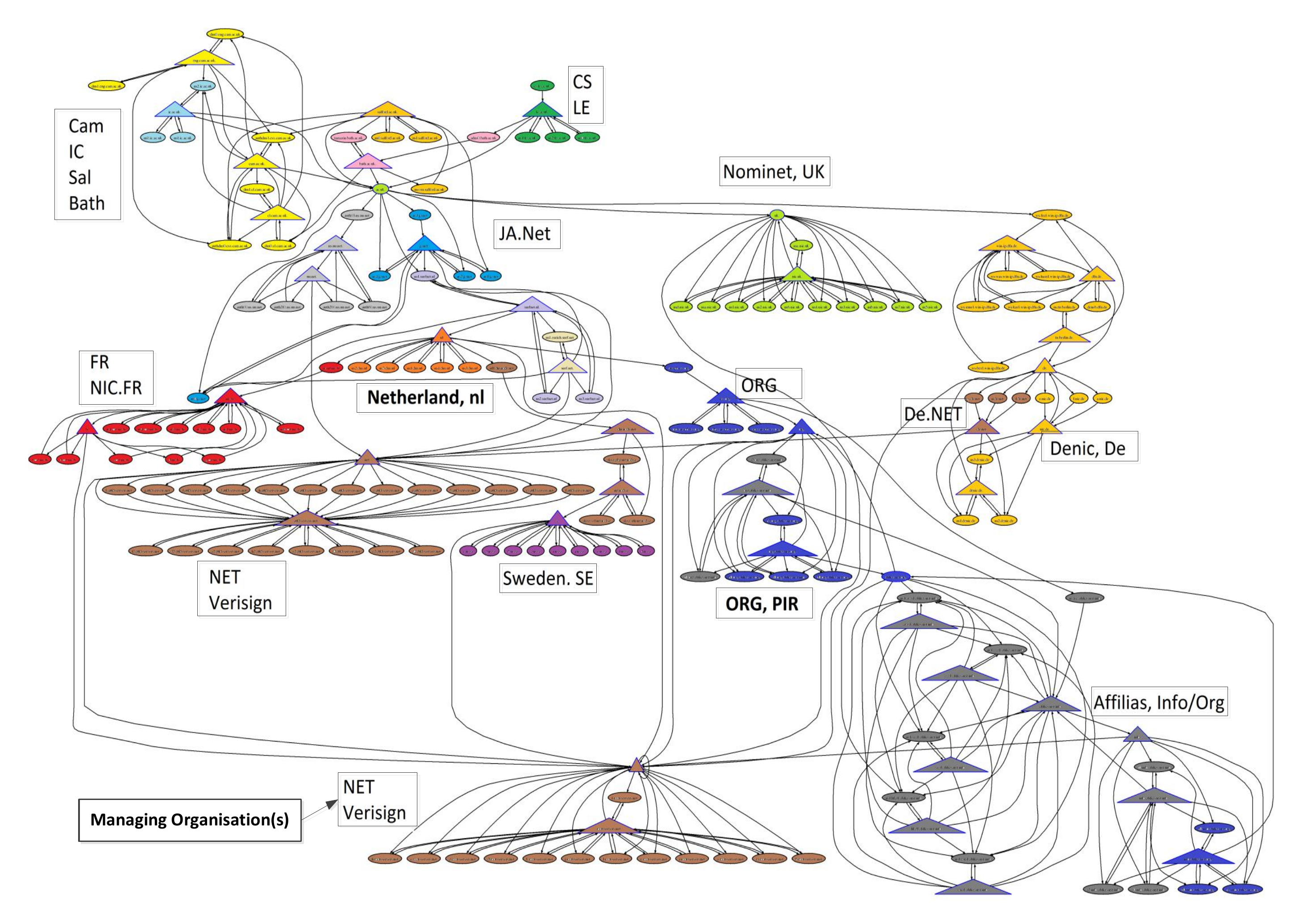}
   \squeezeup
    \squeezeup
 \caption{Name Dependency Graph of (le.ac.uk).}
 \label{fig:f2}
  \squeezeup
   \squeezeup
\end{figure}

Figure~\ref{fig:f2} shows the delegation graph of the zone (le.ac.uk) where the zone le.ac.uk depends on 4 authoritative name servers (ns0, ns1 and ns2.le.ac.uk) under the management of the University of Leicester (UoL), while the fourth name server (adns0.bath.ac.uk) is managed by the University of Bath. In order to resolve any domain under the zone (le.ac.uk), resolver will ask the name servers of the root zone down to the set of authoritative name servers of the zone. While Leicester University directly trusts bath.ac.uk to serve its namespace, it has no control over the name servers that Bath trusts (i.e. name servers under Cambridge, Salford, and Imperial College and so on). Each name server or group of name servers are administered by different organization which creates another layer of transitive trust dependencies amongst those organizations.
\squeezeup
 \squeezeup
\subsection{Operational Planes}
The zone's data plane is the interconnected graph of all infrastructure resource records defined within the zone's configuration file. The interconnected graph of all authoritative name servers involved in the resolution process of a domain within a certain zone is called the zone's control plane and the interconnected graph of all administrative units involved is called the management plane. One reason that the DNS is so powerful is that its data plane allows administrators a great deal of flexibility: they can manage their name space however they like. However, the control and management planes' flexibility can lead to operational problems if not managed conscientiously.
\squeezeup
 \squeezeup
\subsection{Dependency Graphs}
\begin{figure}[ht]
  \centering
  \includegraphics[trim=0cm 2cm 0cm 0cm, clip=true,width=0.7\textwidth]{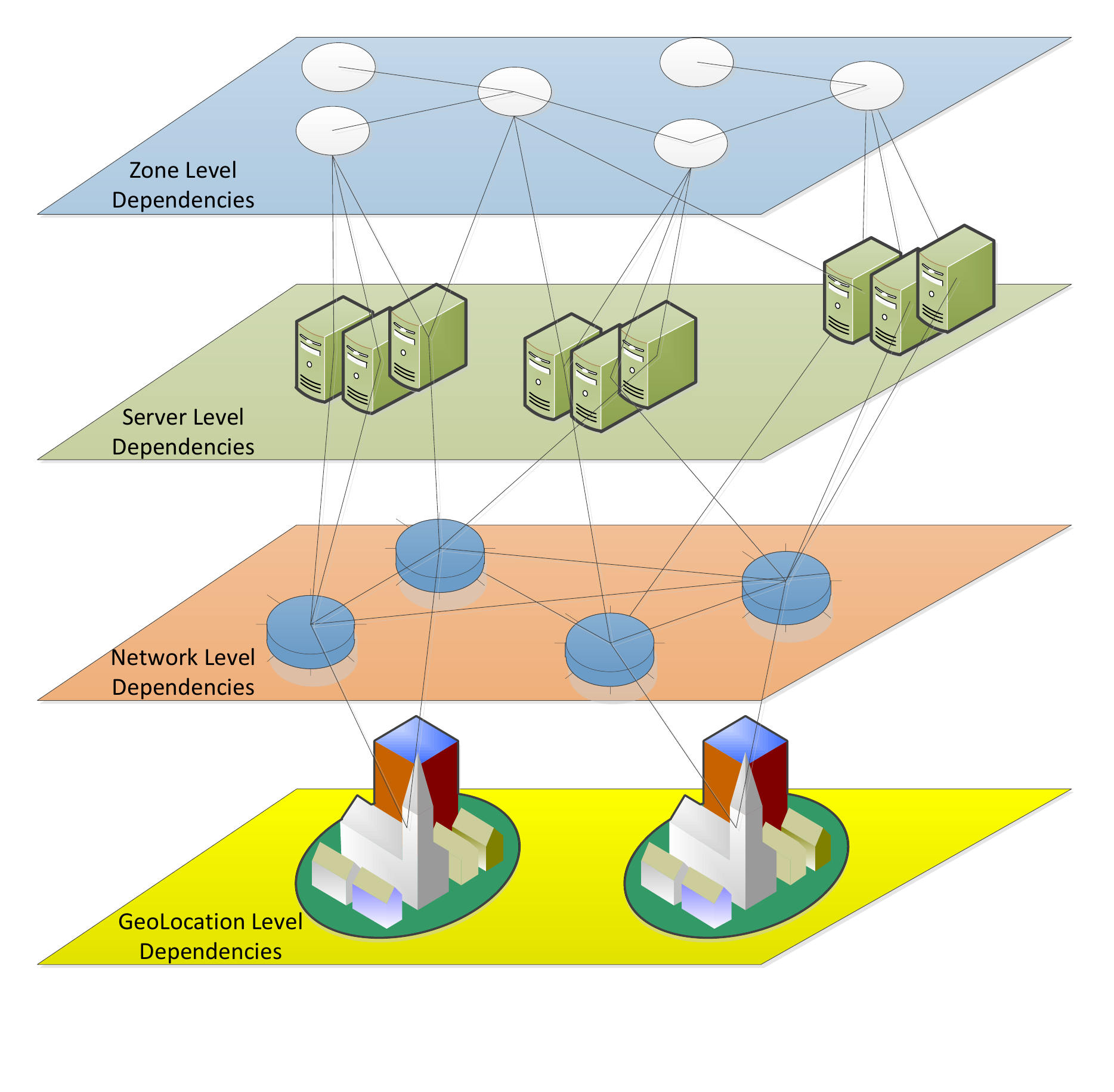}
   \squeezeup
 \caption{DNS Operational Planes and Their Dependencies.}
 \label{fig:f3}
  \squeezeup
  \squeezeup
\end{figure}

The recursive structures of inter-dependencies within and between the DNS operational planes is represented by dependency graph. A dependency graph \cite{deccio2010} is a directed connected graph with a distinguished node (r) which is the root zone. Each node in the graph represents a zone name, and each edge signifies that its source is directly dependent on its target for proper resolution of itself and any descendant domain names. Dependency graphs capture most attributes and relationships between the various operational entities within the DNS and they can be effectively utilized in detecting configuration weaknesses and servers’ deployment problems. Figure~\ref{fig:f3} shows deferent dependencies that occur at the different DNS operational planes.\par

Since many of the misconfigurations can't be detected from the zone file or deployments directly, there is a need for an operational model that encompasses all information related to the zone file and the server deployments in one conceptual graph. The instance of the model (the dependency graph) will enable us to detect zone integrity violations as well as violations in the deployment of name servers and the choice of peering organizations and management structures.  The conceptual graph representation facilitates modelling at multiple levels of details simultaneously.
\squeezeup
 \squeezeup
\section{DNS Operational Model}
\label{sec:sec3}

The DNS Operational Model aims to support operational goals, such as detecting violations of the design and deployment principles, at the authoritative level. To this end we have to search for certain patterns indicating such violations in the instances of the operational model of the system, i.e., the dependency graphs. This means we have to be able to specify a problem as a pattern, and to query the dependency graph about the existence and occurrences of the specified pattern. The model is composed of the following elements:
\begin{itemize}
\item Operational Entities (e.g. resource records, zones, servers and organizations)
\item Properties of operational entities such as (in-bailiwick and out-of-bailiwick name servers)
\item Relations between the entities (e.g. access attributes such as dependability, containment, delegation and management)
\end{itemize}

\begin{figure}[ht]
  \centering
 \includegraphics[trim=1cm 3.2cm 1cm 3.5cm, clip=true, width=0.95\textwidth]{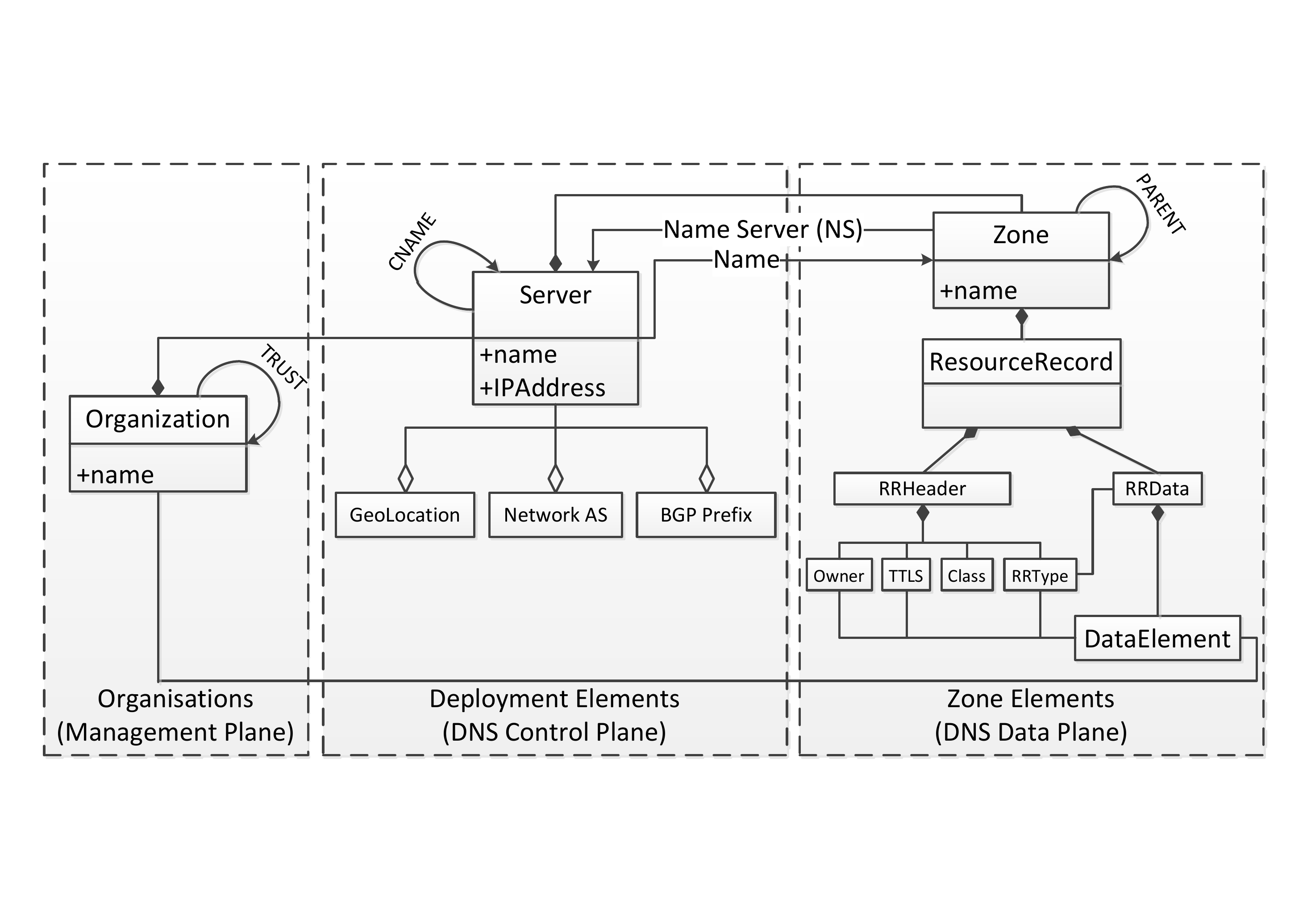}
  \squeezeup
   \squeezeup
 \caption{DNS Operational Model.}
 \label{fig:f4}
  \squeezeup
   \squeezeup
\end{figure}

The operational DNS entities that appear in our model fall into two categories: primitive and composed entities. Composed entities have an identity and a set of properties. In addition to these, composed entities have a list of contained entities, which are primitive or composed entities. A composed entity type is one that contains other entities. The model supports the following composed entities: Organization, Server, Zone and Resource Record. In order to describe a composed entity we have to specify its properties, containment structure (i.e. the entities that it contains), relations and container entity. As an example, we can look at the server component where it can be managed (contained) by organizations. Multiple servers can be managed by one organization. The server can host many zone files and it has the name and IP address as attributes. There are many types of servers and in this context we are concerned with in-bailiwick servers whose name is within the zone file hosted at that particular server and out-bailiwick servers who has a name from a zone hosted in another name server.\par

Three specific dependencies are present within the DNS operational planes and they are:
\begin{enumerate}
\item Parent Dependency: resolving the name of a domain name is always dependent on resolving its parent name since the resolver must learn the authoritative servers for a zone from referrals from the zone’s hierarchical parent.
\item Authoritative Name Server (NS) Dependency: A zone is said to depend on a name server if the name server could be involved in the resolution of names in that zone.
\item CNAME Aliasing Dependency (Name pointing to another Name): the resolution of an alias is always dependent on the resolution of its target CNAME. If a resolver receives a response indicating that the name in question is an alias to another name, it must subsequently resolve the target of the alias, and so on until an address is returned.
\end{enumerate}

The dependency graph can be extracted from the zone file and from the chain of authoritative name servers and organizations involved in the resolving process of domains under that particular zone. This is done by analysing the zone file and the dependencies between the different resource records and their data elements and by following the query process as outlined in Fig.~\ref{fig:f1} using certain DNS tracing tools extensively. All types of dependencies and recursive queries are followed to get the full dependency graph of the zone in the three operational planes.
\squeezeup
\squeezeup
\section {Operational Bad Smells}
 \squeezeup
\label{sec:sec4}
In software engineering, bad smells in code \cite{fowler1999} identify risks to non-functional quality in a software system based on structural properties and metrics. We transfer these ideas to the realm of the DNS, where operational bad smells are configuration and deployment choices by zone administrators that are not errant or technically incorrect, and do not currently prevent the system from doing its designated functionality. Instead, they indicate weaknesses that may impose additional overhead on DNS queries, or increase the system vulnerability to threats, or increase the risk of failures in the future.\par
The set of identified bad smells is being formally specified in concise and reusable terms based on a template that includes the bad smell name, type, inspection plane(s), description \& occurrences, quality impacts and detection strategies. The catalogue will be expanded by including refactoring rules for each smell and how these rules have to be applied on the model instance to eliminate the concerned bad smell. Examples of catalogue entries are shown in Table~\ref{tab:t2} and Table~\ref{tab:t4}  listed as part of the case studies in Section~\ref{sec:sec5}.\par
Although DNS troubleshooting techniques and problem identification methods have been proposed and several tools have been built, most of these methods and tools apply their detection techniques directly on the zone files through a predefined zone schema and integrity constraints. They don't take into account the inter-dependencies stemming from the hierarchical nature of the DNS or the zone administrators practices. Instead, we propose a model-based approach that subsumes all the steps necessary to identify, specify and detect the DNS operational bad smells. The \textbf{\textit{ISDR}} method is composed of four stages and produces the operational bad smells catalogue:
\begin{enumerate}
\item \textbf{I}dentification, including domain analysis using DNS standards in the form of Request for Comments (RFCs), best practices and policy documents, literature review and DNS expert views.
\item \textbf{S}pecification of a set of operational bad smells using a reusable vocabulary and classification of the bad smells in a taxonomy that shows the scope of the inspection element or plane and system's external qualities affected by the smell.
\item \textbf{D}etection of bad smells in the form of general detection queries and formulas.
\item \textbf{R}efactoring as a correction mechanism to the operational bad smells. Other correction mechanisms may be formulated in the form of reports or reconfiguration recommendations.
\end{enumerate}
The following are the ISDR method stages in details:
\squeezeup
\subsection {Identification}

The first stage in our method consists of performing deep analysis of the DNS standards, Request for Comments (RFCs), best practices and policy documents to identify weaknesses in configuration and deployment choices made by administrators that may impose additional overhead on DNS queries, or increase the system vulnerability to threats, or increase the risk of cascaded failures.
\squeezeup
\subsection{Specification}

The weaknesses identified in the previous step,  termed as \textit{operational bad smells}, are then defined using certain key terms, unified vocabulary and reusable concepts in this domain. We developed a taxonomy that describes the structural relationships between the various bad smells. The taxonomy has an important role in defining the scope of inspection and highlighting the metrics or structural properties related to the bad smell. It classifies the bad smells based on the following categories:
\begin{enumerate}
\item	Operational plane: Data, control and management planes.
\item	Affected entity types: Single type, inter-type, intra-type, or inter-zone.
\item	Property of the smell: Lexical, structural or measurable.
\end{enumerate}

\begin{figure}[ht]
  \centering
 \includegraphics[trim=3.5cm 0.5cm 3.5cm 0.5cm, clip=true,width=0.75\textwidth]{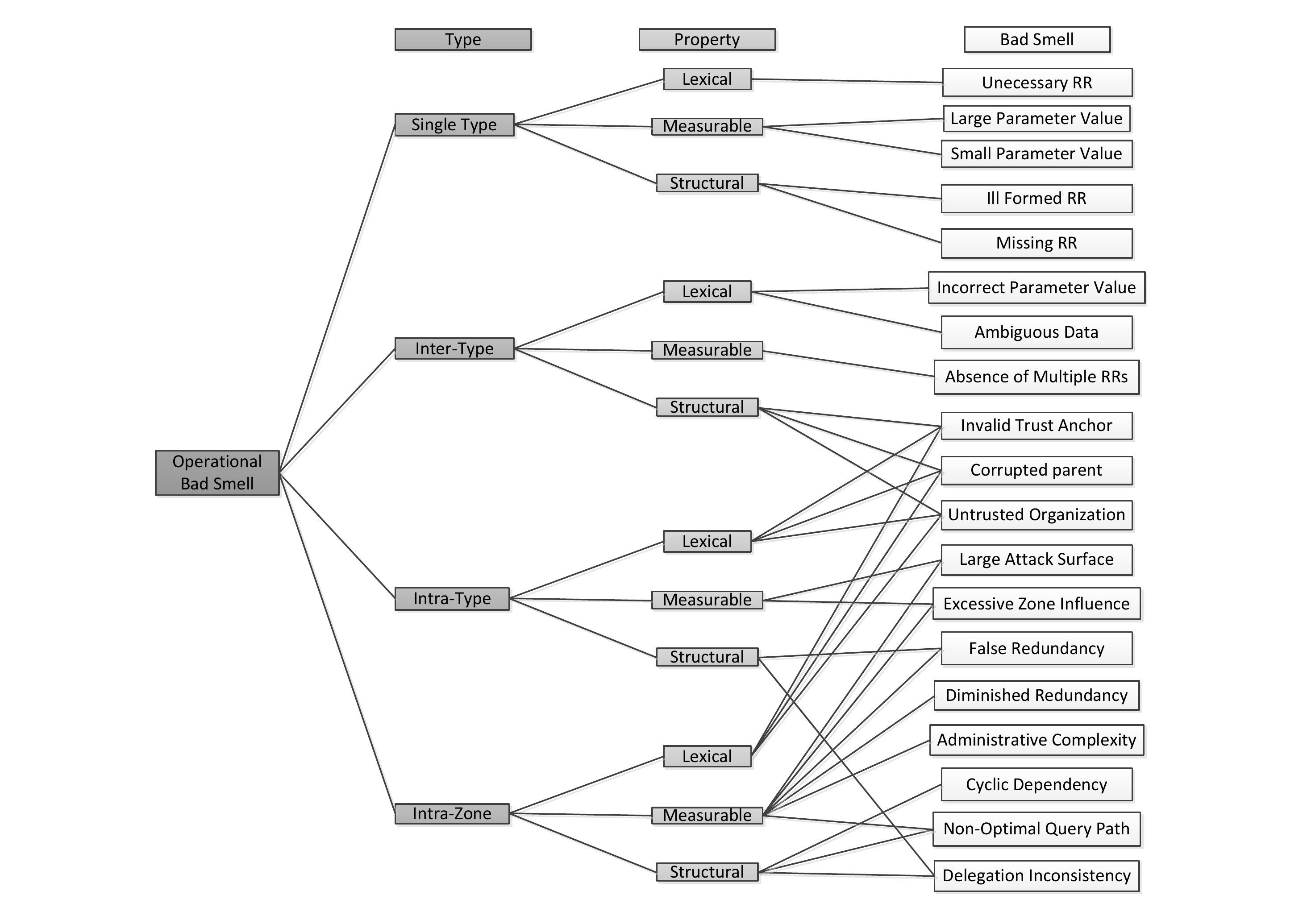}
  \squeezeup
 \caption{DNS Operational Bad Smells Taxonomy.}
 \label{fig:f5}
  \squeezeup
   \squeezeup
\end{figure}

Figure~\ref{fig:f5} shows a partial graphical representation of the DNS operational bad smells taxonomy. The taxonomy is generic and defines a bad smell in more than one category. It can easily be extended by defining new categories of bad smells based on subsequent iterations of the DNS operational domain analysis. So far we have already identified 19 bad smells that can be used as a representative set that spans the different operational planes with various detection properties.\par

In the context of metrics-based analysis techniques, the aforementioned classification of design entities (as explained in Section~\ref{sec:sec3}) has a particular relevance: it provides a pertinent explanation about why metrics are defined and computed only for some entity types (i.e., organizations, network, server, zones and resource records). The explanation resides basically in the distinctive aspects that exist between the two, i.e., the fact that a composed entity can contain other entities and that it can have relations with other entities.  As direct measurements are mainly ”counting” the different entities contained in, or related to a measured entity, it becomes obvious why the object of measurement is restricted to composed design entities. Interesting examples of metrics are per-server and per-zone distributions such as:
\squeezeup
\begin{enumerate}
\item The number of zones that are served from multiple name servers in different network autonomous systems or diverse geographical locations \textit{(Server Redundancy)}.
\item The number of zones that influence the resolving of domain names within a particular zone \textit{(Zone Influence)}.
    \squeezeup
\end{enumerate}

For the proper interpretation of each structural metric defined over the operational model, we give the metric definition, usability, how to measure and a formula for computing that metric. Table~\ref{tab:t0} shows the interpretation model for the metric \textit{Administrative Complexity} \cite{deccio2010}. \par

\begin{table}[ht]
\centering
\squeezeup
\caption{Interpretation of the Administrative Complexity Metric.}
 \label{tab:t0}
\begin{tabular}{ |  p{2.6cm} | p{12cm} |}
   \hline
     Metric &Administrative Complexity. \\
\hline
  Definition &Describes the diversity of a zone with respect to the organisations administering its authoritative name servers. \\
\hline
Usability &The advantage mutual hosting of zones between organizations is an increased availability but at the same time increased potential of failure and instability of the zone resolution process. \\
\hline
 How to Measure & Count the number of authoritative name servers of each organization involved in the dependency graph of zone(z). \\
\hline
    Metric Notation &$O_z$: set of organizations administering authoritative name servers hosting zone ($z$); $n$: total number of authoritative name servers of zone ($z$); $NS^o_z$: the subset of name servers administered by organization $o$ in $O_z$. \\
\hline
    Formula &$Ac(z)=$1-
$\sum_o^n$
$(\frac{NS^o_z}{NS_z})^n$. \\
\hline
\end {tabular}
\squeezeup
\end{table}
\squeezeup
\squeezeup
\subsection {Detection}
In order to be able to detect bad smells occurring on model instances, we need to capture deviations of the particular instance model from the good and recommended operational best practices. Lexical and structural properties are used to detect some of the bad smells using direct queries on the instance model such as (Are there any cycles in the dependency graph.?).The metric-based approach combines a set of metrics and set operators to compare them against absolute or relative threshold values. Setting the absolute or relative operational metrics threshold values can be done using local policy constraints or best practices from the wider DNS domain literature and expert views.
\squeezeup
\subsection{Refactoring}
In the area of object-oriented programming, refactoring \cite{opdyke1992} is the technique of choice for improving the structure of existing code without changing its external behaviour. Graph-based, general refactoring rules \cite{bisztray2009} will be suggested to remove the bad smells identified and detected in the previous stages. The general approach of refactoring \cite{mens} is to include the following steps: (1) identify the location for refactoring, (2) determine which refactoring rules should be applied and on what sequence, (3) guarantee that refactoring rules are preserving the external behaviour of the system, (4) application of selected refactoring rules, (5) assess the effect of refactoring on the system’s external qualities and (6) maintain the consistency between the refactored elements and other system artefacts.
\squeezeup
\subsection {Method Execution and Tool Support}
The ISDR method is executed on a particular instance of the DNS operational model (Dependency Graph) using the following steps:
\begin{itemize}
\item\textbf{Step 1:} Extract the dependency graph from the zone configuration file and the authoritative servers' deployment.
\item\textbf{Step 2:} Query the dependency graph for any bad smell using the methods and metrics defined in the Bad Smells Catalogue.
\item\textbf{Step 3:} Apply relevant refactoring rule(s) on all matching occurrences of the LHS of the rule on the instance model. A new dependency graph is generated as an output of this step.
\item\textbf{Step 4:} New zone file(s) and authoritative name servers deployment layout can be automatically generated from the new Dependency Graph or a set of recommendations can be presented to the system administrator with relevant quality impacts.
\end{itemize}
The method will be implemented using a pre-emptive diagnostic advisory tool that will detect and  flag configuration changes that might decrease the robustness, resilience or security posture of a domain name, before even the changes become into production.
\squeezeup
\squeezeup
\section{Validation}
\label{sec:sec5}
We validate our method by applying it and its associated execution technique to several bad smells where some of them have been already identified as misconfigurations in the literature \cite{pappas2009,deccio2010,kalafut2008,herzberg2013,lu2014}.
\squeezeup
\subsection{Case Study (1): Cyclic Dependency}
To achieve acceptable geographical and network diversity, zone administrators often establish mutual arrangement with peer organizations to host each other’s zone files. Authoritative name servers located in other zones are normally identified by their names instead of their addresses and called out-of-bailiwick name servers. A cyclic zone dependency \cite{pappas2009} occurs when two or more zones depend on each other in a circular way.\par

Table~\ref{tab:t1} shows that the zone (example.com) has 4 authoritative name servers responsible for resolving domain names under this zone as defined in its parent zone (.com). Two servers (ns1 and ns2.example.com) are \textit{in-bailiwick} servers and it is mandatory to include their IP addresses in the parent zone in order to properly resolve domain names under that zone. The other two servers (dns1 and dns2.example.net) are located in another zone and there is no need to include their IP addresses in the (.com), example.com parent zone file. On the other hand, the (.net) zone which is the parent of the (example.net) zone, is served by two \textit{out-of-bailiwick} name servers located in the (example.com) zone.\par

\begin{table}[ht]
\squeezeup
\squeezeup
\caption{Content of Zone File for Case Study (1).}
 \label{tab:t1}
\begin{tabular}{| p{7cm} | p{0.5cm} | p{7cm} |}
 \cline{1-1}
\cline{3-3}
\$ORIGIN .com. && \$ORIGIN .net. \\
 \cline{1-1}
\cline{3-3}
\end {tabular}
\begin{tabular}{ |  p{2.8cm} | p{0.55cm} | p{2.8cm} | p{0.52cm} | p{2.8cm} | p{0.55cm} |p{2.8cm} |}
   example.com. &NS & ns1.example.com.&&example.net.&NS&ns1.example.com. \\
     example.com. &NS & ns2.example.com.&&example.net.&NS&ns2.example.com. \\
\cline{5-7}
     example.com. &NS & dns1.example.net.& \multicolumn{3}{c}{} \\
     example.com. &NS & dns2.example.net.& \multicolumn{3}{c}{} \\
 \cline{1-3}
     ns1.example.com. &A& 1.1.1.1& \multicolumn{3}{c}{} \\
     ns2.example.com. &A & 1.1.1.2& \multicolumn{3}{c}{} \\
 \cline{1-3}
 \end{tabular}
\squeezeup
\end{table}

In this example, the two zones work nicely under normal circumstances but if (for any reason), both in-bailiwick name servers become unavailable, both example.com and example.net zones will not be reachable because the IP Addresses of the other two authoritative name servers can't be resolved. This example illustrates the failure dependency between zones, where the failure of some servers in one zone will render the other zone unreachable. The quality impacts of such a bad smell are significant reduction on availability and resiliency of the zone against multiple points of failure.

\begin{figure}[ht]
\squeezeup
  \centering
   \includegraphics[trim=1.3cm 1.7cm 1.8cm 1cm, clip=true,width=0.75\textwidth]{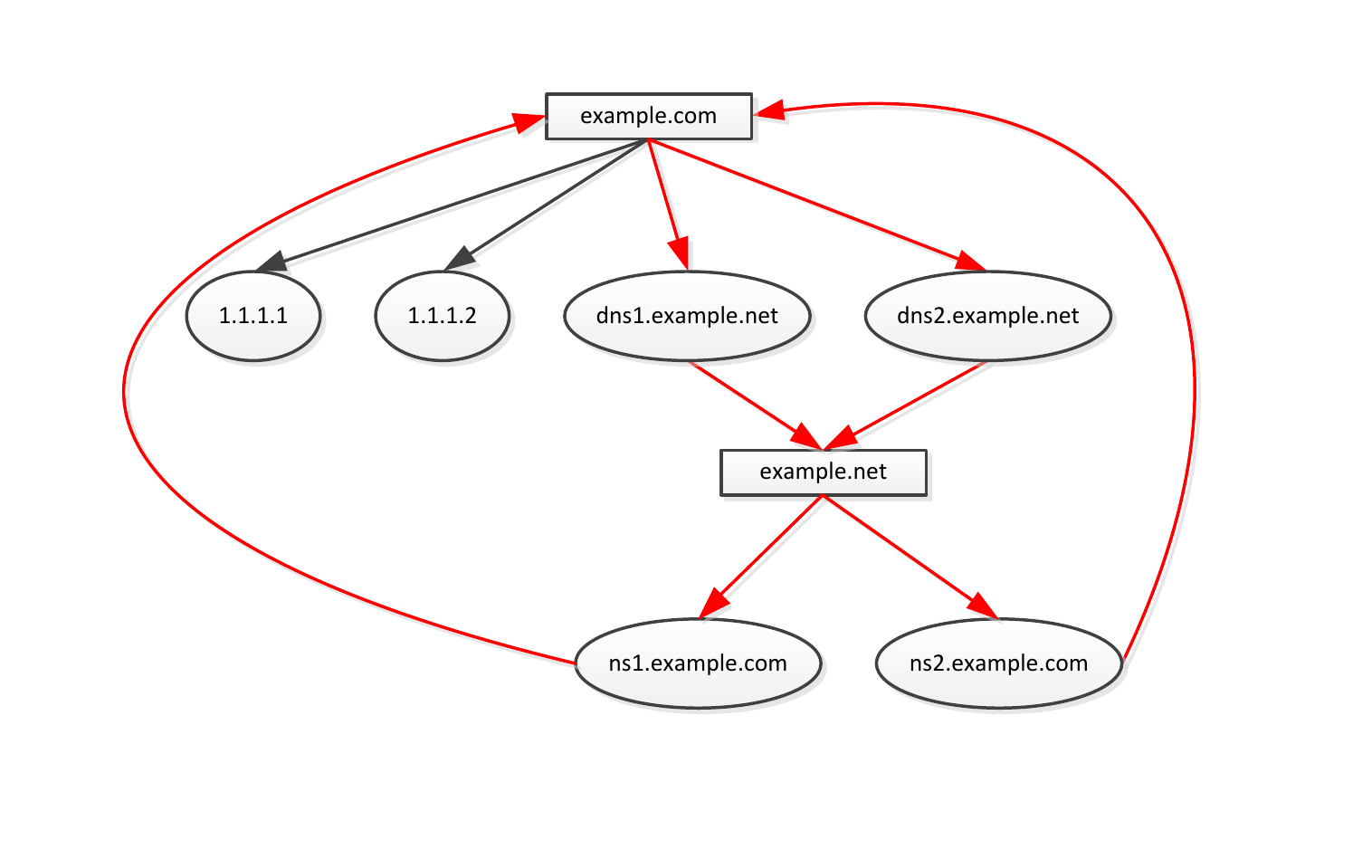}
 \caption{Part of the Dependency Graph of Case Study (1).}
 \label{fig:f7}
 \squeezeup
\end{figure}

Checking each zone individually for configuration errors will not lead to the detection of this bad smell since they are both configured correctly. On the other hand, constructing the dependency graph will easily show the occurrence of two circular paths that identify the smell. \par

\begin{table}[ht]
\centering
\caption{Catalogue Entry for the Cyclic Dependency Bad Smell.}
\squeezeup
 \label{tab:t2}
\begin{tabular}{ |  p{3cm} | p{9cm} |}
   \hline
     Name &Cyclic Dependency. \\
\hline
  Type &Intra-Zone, Structural. \\
\hline
Inspection Planes &Data and Control Planes. \\
\hline
 Occurrences & Cyclic zone dependency occurs when two or more zones depend on each other in a circular way. \\
\hline
    Quality Impacts &Reduced availability and reduced resiliency. \\
\hline
    Detection Strategy &Is there any cycle in the Dependency Graph? (Query on the DNS Operational Model Instance). \\
\hline
   Correction Mechanism (Refactoring) &Add a glue record for the (out-of-bailiwick) authoritative name servers involved in the cycle in the zone file. \\
\hline
\end {tabular}
\end{table}

\begin{figure}[ht]
  \centering
   \includegraphics[trim=1cm 0.5cm 0.5cm 0.5cm, clip=true, width=0.75\textwidth]{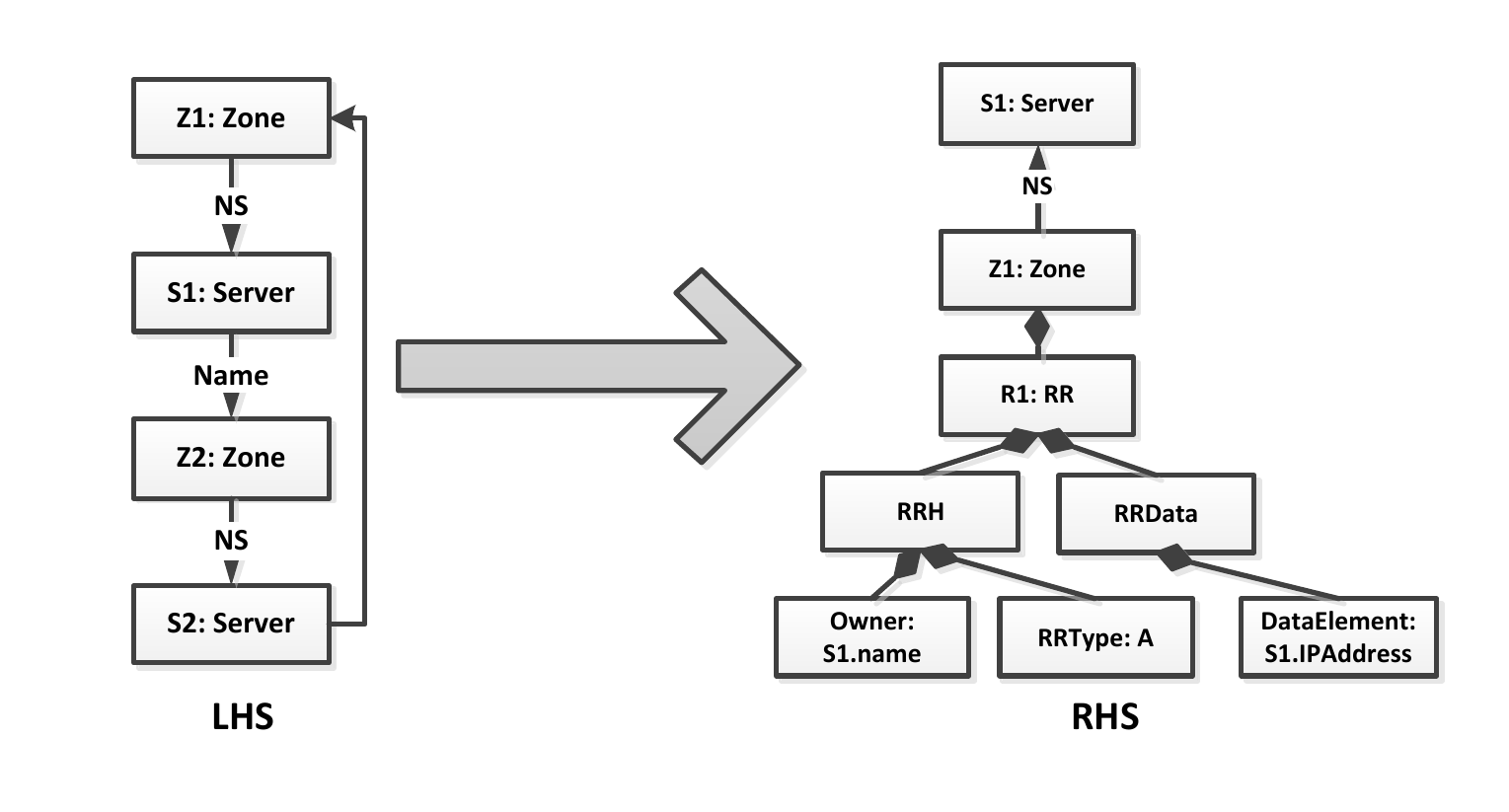}
\squeezeup
\squeezeup
 \caption{Refactoring Rule: AddGlueRecord.}
 \label{fig:f8}
\end{figure}

Cyclic Dependencies can be eliminated by the inclusion of specific resource records (RRType: A) for both out-of-bailiwick servers in the (.com) zone. This will enable resolving the domain names under the (example.com) and (example.net) zones even when the two in-bailiwick servers are unreachable. We execute this correction mechanism in the form of a graph transformation based refactoring rule (AddGlueRecord) applied on the dependency graph as shown in Figure~\ref{fig:f8}. Since we have two matches for the LHS of the rule on the actual instantiation of the model (the dependency graph in Figure~\ref{fig:f7}), then the rule needs to be applied twice in order to remedy all occurrences of the bad smell. A new zone file can then be automatically generated from the new Dependency Graph as shown in Table~\ref{tab:t3} or a set of recommendations can be presented to the system administrator to eliminate the bad smell.\\
\squeezeup
\squeezeup

\begin{table}[ht]
\squeezeup
\caption{New Zone File Generated After Executing the Refactoring Rule(s).}
\squeezeup
 \label{tab:t3}
\begin{tabular}{| p{7cm} | p{0.5cm} | p{7cm} |}
 \cline{1-1}
\cline{3-3}
\$ORIGIN .com. && \$ORIGIN .net. \\
 \cline{1-1}
\cline{3-3}
\end {tabular}
\begin{tabular}{ |  p{2.8cm} | p{0.55cm} | p{2.8cm} | p{0.52cm} | p{2.8cm} | p{0.55cm} |p{2.8cm} |}
   example.com. &NS & ns1.example.com.&&example.net.&NS&ns1.example.com. \\
     example.com. &NS & ns2.example.com.&&example.net.&NS&ns2.example.com. \\
\cline{5-7}
     example.com. &NS & dns1.example.net.& \multicolumn{3}{c}{} \\
     example.com. &NS & dns2.example.net.& \multicolumn{3}{c}{} \\
 \cline{1-3}
     ns1.example.com. &A& 1.1.1.1& \multicolumn{3}{c}{} \\
     ns2.example.com. &A & 1.1.1.2& \multicolumn{3}{c}{} \\
     dns1.example.net. &A & 1.1.1.3& \multicolumn{3}{c}{} \\
     dns2.example.net. &A & 1.1.1.4& \multicolumn{3}{c}{} \\
 \cline{1-3}
 \end{tabular}
 \squeezeup
 \squeezeup
\end{table}

\squeezeup
\subsection{Case Study (2): False Redundancy}

Redundancy \cite{rfc2182} is one of two mechanisms used by DNS administrators to ensure high availability of domain names. The level of availability provided by redundant servers is a function not only of their number, but also of their physical location and the networks they connect to.  In 2001, a DNS bad deployment choice \cite{microsoft} caused many Microsoft's web sites and email servers to be unreachable (although they were actually still operational). All authoritative name servers for the zone (microsoft.com) were place in one location, connected to the same network, and placed behind one particular network router. When the router failed, this local bad choice has a large global impact by increasing the queries on one of the DNS root servers (F server) from the normal 0.003\% of all queries to over 25\% \cite{pappas2009}. The catalogue entry for the False Redundancy bad smell is shown in Table~\ref{tab:t4}.\par

\begin{table}[ht]
\centering
\squeezeup
\caption{Catalogue Entry for the False Redundancy Bad Smell.}
 \label{tab:t4}
\begin{tabular}{ |  p{3cm} | p{11.5cm} |}
   \hline
     Name &False-Redundancy. \\
\hline
  Type & Measurable and Inter-zone. \\
\hline
     Inspection Planes &Control Plane. \\
\hline
   Occurrences &  When all redundant servers are located within the same physical location, connected to the same network, placed within the same address prefix. \\
\hline
   Quality Impacts & Reduced availability, decreased resilience, and the system become susceptible to single point of failure at certain granularity. \\
\hline
   Detection & Queries on the dependency graph regarding the following metrics: a) number of authoritative name servers, b) geographical locations servers are placed in, c) networks connected to, and d) BGP prefixes, \\
\hline
   Refactoring & Applying the MoveServerLocation refactoring rule will ensure the availability of the zone and its resilience to a single point of failure.  \\
\hline
 \end{tabular}
 \squeezeup

\end{table}

\begin{figure}[ht]

  \centering
  \includegraphics[trim=0.65cm 0.7cm 0.68cm 0.65cm, clip=true,width=0.75\textwidth]{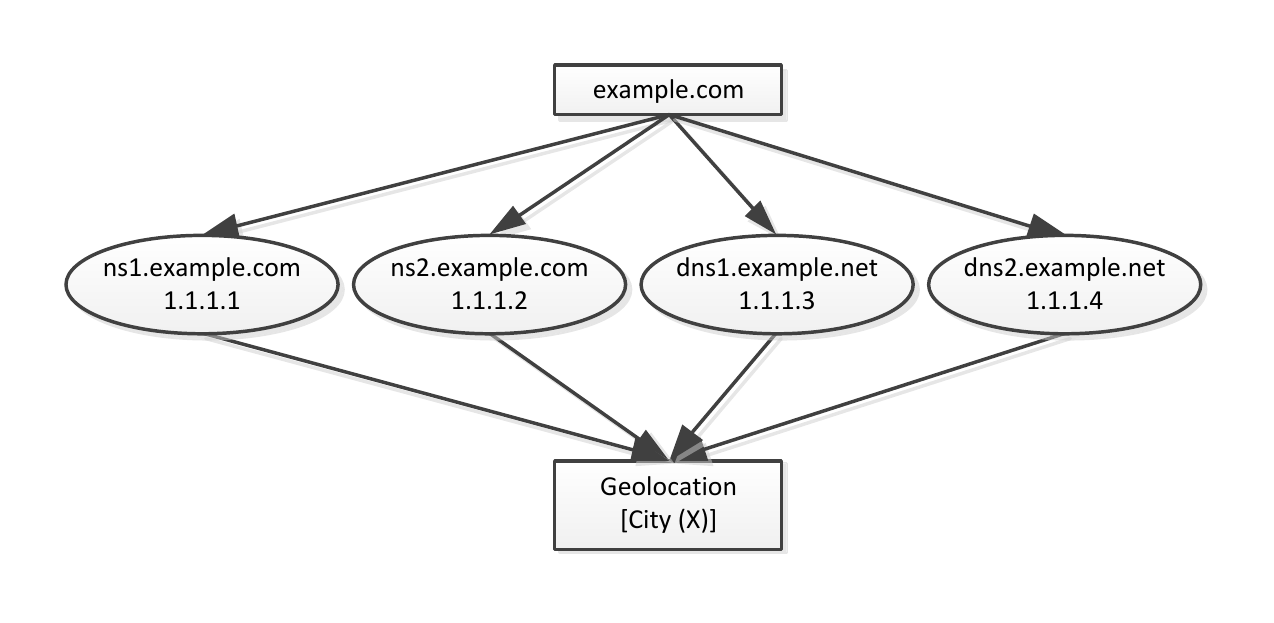}
  \squeezeup
 \caption{Geographical Location Dependency Graph of Case Study (2).}
 \label{fig:f9}
\squeezeup
\end{figure}

In this example, we are looking into one aspect of False Redundancy, which is the geographical placement of the authoritative name servers. Looking at the dependency graph extracted from the zone file, generated as an output of case study (1) and the deployment of authoritative name servers, as shown in Figure ~\ref{fig:f9}, it is clear that the \textit{geographical redundancy} of the zone (example.com) is one which is much less than the server's Redundancy which is supposed to be 4 (the total number of authoritative name servers defined in the zone). Looking at the IP address associated with each of these servers, it is evident that all of them are connected to the same network, and behind even the same router. This deployment choice introduces a single point of failure since all servers are located in the same geographical area and this badly affect the resiliency and availability of the zone and its domain names. Geographical area may be defined as a continent, country, city or even a certain building, which may also be susceptible to power outage, natural disaster or any other risk.

In order to detect the occurrence of the False Redundancy bad smell, one set of queries regarding the number of authoritative name servers of the particular zone, number of distinct geographical locations where those servers are placed in, can be executed against the dependency graph in Figure~\ref{fig:f9}. The resulted measurements are used in detecting the bad smell as defined in its catalogue entry. The threshold values for the metrics are set based on the best practices and policies as identified in the first step of the ISDR method or can be left to the system administrator to set based on the local policies and operational requirements.\par

\begin{figure}[ht]
  \centering
   \includegraphics[trim=0cm 0.7cm 0cm 0.4cm, clip=true, width=0.75\textwidth]{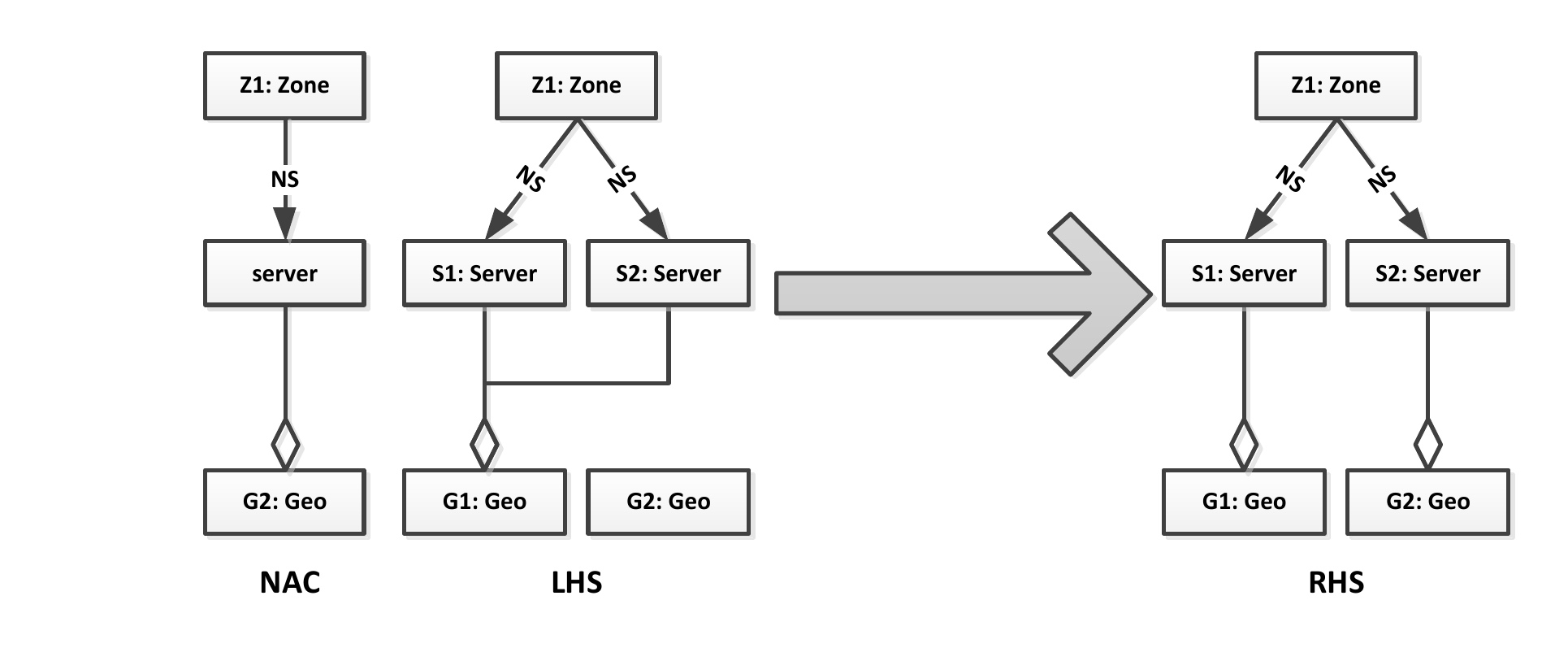}
    \squeezeup
 \caption{Refactoring Rule: MoveServerLocation.}
 \label{fig:f11}
  \squeezeup
\end{figure}
 \squeezeup
It should be noted that the refactoring rule shown in Figure~\ref{fig:f11} is just one of the options to eliminate this bad smell, i.e., there could be another rule for creating a new server in a new location rather than moving an existing one. We can look at network number or BGP prefix instead of location. It can also take more than one rule application to resolve the situation, so a single rule specifies an incremental improvement, which may have to be repeated or combined with others.
\squeezeup
\section{Related Work}
\label{sec:sec6}
\subsection{DNS Interdependencies and Misconfigurations}
Ramasubramanian, et al. \cite{ram2005} demonstrate the far-reaching effects of DNS dependencies. Their results show that a domain name relies on 44 name servers on average. Deccio et al. \cite{deccio2010} perform further examination of name resolution behaviors to create a formal model of name dependencies in the DNS and quantify the significance of such dependencies. Several surveys of production DNS deployments have been conducted \cite{pappas2009,kalafut2008,wessels2004} with various misconfigurations are analyzed. So far the main efforts in addressing the problem have focused on informing the operators about the existence of DNS configuration errors, either by Internet RFCs \cite{rfc1912,rfc2182} or with directives set by specific organizations \cite{wg4}.
\squeezeup
\squeezeup
\subsection{DNS Troubleshooting}
\squeezeup
Although several DNS troubleshooting techniques and problem identification methods \cite{Pappas2004tool,decciodnsviz} have been proposed and several tools \cite{dnschecker,intodns,zonemaster} have been built, most of these methods and tools apply their detection techniques directly on the zone files through a predefined zone schema and a specified set of integrity constraints. Chandramouli and Rose \cite{chan} considered integrity constraints for Resource Records (RRs) from single and multiple zones. They found that many integrity constraints have to be satisfied across zones. \cite{casalicchio2012} proposed a set of metrics to be used to evaluate the health of the DNS by measuring the DNS along three dimensions, namely vulnerability, security and resilience. Most of these studies can detect only a subset of the DNS infrastructure-related configuration errors. On the other hand, they implement diagnostic tests that can identify errors related with application-specific resource records. They do not take into account the inter-dependencies stemming from the hierarchical nature of the DNS, zone administrators' practices and deployment choices.\par

Despite all the existing efforts, DNS configuration errors are still widespread today \cite{lu2014}, and one of the main reasons is the lack of adequate tools to help DNS operator identify and correct configuration errors in their own domains. Previous studies are largely based on empirical analysis, whereas in this paper we derive a formal operational model and methodology to systematically identify misconfigurations and bad deployment choices in the form of operational bad smells.
\squeezeup
\squeezeup
\subsection{Bad Smells and Refactoring}
There is a large body of work on the identification of problems in software, database and networks. Several books have been written on smells \cite{fowler1999} in the context of object-oriented programming. Marinescu \cite{mari} presented a metric-based approach to detect code smells. Alikacem and Sahrouri \cite{alikacem} proposed a language to detect violations of quality principles and smells in object-oriented systems. Mens and Tourwe \cite{mens} have conducted a comprehensive survey of software refactoring. While software refactoring has started focusing on restructuring of programs, the research has extended to model refactoring \cite{emf}. \par

Our objectives are similar to those of previous DNS operation studies but our approach differs. Our method utilizes a set of measurable, structural and lexical properties defined over a DNS operational model to detect the smells in early stages of the DNS deployment. It also suggests graph-based refactoring rules as correction mechanisms for those bad smells.

\squeezeup
\squeezeup
\section {Conclusion and Future Work}
 \squeezeup
\label{sec:sec7}
Currently, there is little consensus on the right measures and acceptable performance levels for the DNS as a whole related to availability, security, stability and resiliency. Individual operators and independent researchers have measured various aspects of the DNS, but to date little progress has been made in defining and implementing standard, system-wide metrics or acceptable service levels. Efforts to improve risk management related to DNS security, stability and resiliency must be guided by an improved ability to measure these characteristics and assess the utility of programs and resources investments. A key enabler of improving this situation will be to ensure that composite parts of DNS operations are correctly configured, deployed, instrumented and measured.\par

The method presented in this paper will lay the basis for developing a visual advisory tool for system administrators to analyse, discover, and remedy operational bad smells. The diagnostic tool will consider several properties and metrics from the DNS operational model presented in this research in relation to the domain name whose zone is being modified. The tool, in a systematic process, can automatically direct the zone administrator to places in the zone file that contain potential design and deployment problems that may compromise availability, resiliency or security of a domain name before the changes become into production. Zone administrator will be able to run several scenarios and apply several refactoring rules through the tool to determine the solution that best meets their local policies.\par

The tool is being designed to cope with zones with very large size and need to be fast enough to be practically applicable. A set of consistent refactoring steps will be applied (or recommended) as graph transformation rules using available tools and techniques. The rule-based behaviour preservation \cite{bisztray2009} will be verified to make sure the suggested rules preserve the system functionality and improve its external qualities. Execution of the refactoring rules may introduce complex sequence of operations to transform the model changes into physical resources relocation. In order to implement some of these refactoring rules, we need to take into consideration access control permissions and physical access to, or coordination actions such as service level agreements (SLAs), with external sites. These concerns will be tackled as part of the refactoring execution steps and available techniques and tools \cite{henshin} are being currently investigated.
\squeezeup
\squeezeup
\squeezeup
\nocite{*}
\bibliographystyle{eptcs}
\bibliography{generic}
\end{document}